\documentclass[runningheads]{llncs}
\usepackage{graphicx}
\usepackage{booktabs}
\usepackage{cite}
\usepackage{subfig}
\usepackage{makecell}

\begin{document}
\title{Distinguishing Healthy Ageing from Dementia: a Biomechanical Simulation of Brain Atrophy using Deep Networks}
\titlerunning{Biomechanical Simulation of Brain Atrophy using Deep Networks}

\author{Mariana Da Silva\inst{1} \and
 Carole H. Sudre\inst{2,3,1} \and Kara Garcia\inst{4} \and Cher Bass\inst{1,5} \and M. Jorge Cardoso\inst{1} \and
 Emma C. Robinson\inst{1}}
 
%1{Da Silva, Mariana}
%2{Sudre, Carole H.}
%3{Garcia, Kara}
%4{Bass, Cher}
%5{Cardoso, M. Jorge}
%6{Robinson, Emma C.}

\authorrunning{M. Da Silva et al.}

\institute{School of Biomedical Engineering and Imaging Sciences, King's College London \and MRC Unit for Lifelong Health and Ageing at UCL, University College London \and Centre for Medical Image Computing, Department of Computer Science, University College London \and Department of Radiology and Imaging Sciences, School of Medicine, Indiana University \and Panakeia Technologies, London, UK}

%=
\maketitle              % typeset the header of the contribution
\begin{abstract}

Biomechanical modeling of tissue deformation can be used to simulate different scenarios of longitudinal brain evolution. In this work, we present a deep learning framework for hyper-elastic strain modelling of brain atrophy, during healthy ageing and in Alzheimer's Disease. The framework directly models the effects of age, disease status, and scan interval to regress regional patterns of atrophy, from which a strain-based model estimates deformations. This model is trained and validated using 3D structural magnetic resonance imaging data from the ADNI cohort. Results show that the framework can estimate realistic deformations, following the known course of Alzheimer's disease, that clearly differentiate between healthy and demented patterns of ageing. This suggests the framework has potential to be incorporated into explainable models of disease, for the exploration of interventions and counterfactual examples.

\keywords{Deep Learning  \and Biomechanical Modelling \and Neurodegeneration \and Disease Progression.}
\end{abstract}

\section{Introduction}

Alzheimer's Disease (AD) is neurodegenerative condition characterized by progressive and irreversible death of neurons, which manifests macroscopically on structural magnetic resonance images (MRI) as progressive tissue loss or atrophy. While, cross-sectionally the progression of the disease is well documented - presenting with disproportionate atrophy of the hippocampus, medial temporal, and posterior temporoparietal cortices \cite{rabinovici2008distinct,carmichael2013coevolution}, relative to age matched controls - in reality disease progression is heterogeneous across individuals and may be categorised into subtypes \cite{ferreira_2017}. Historically, this has meant that early stage AD has been challenging to diagnose from structural MRI changes alone \cite{frisoni2010clinical, hongming_dl_ad_2019, bae2020identification, khan2019transfer}. 

Biomechanical models present an alternate avenue, in which rather than performing post-hoc diagnosis of AD from longitudinally acquired data, it instead becomes possible to build a forward model of disease, simulating different possible scenarios for progression  \cite{khanal_biophysical_2014, khanal_biophysical_2016}. Such models have been used broadly throughout the literature to simulate both atrophy and growth \cite{richman_mechanical_1975, xu_axons_2010, tallinen_growth_2016} and are usually based on hyperelastic strain models, implemented using finite element methods (FEM) \cite{tallinen_gyrification_2014} or finite difference methods (FDM) \cite{khanal_biophysical_2014}. 
 
Accordingly, in this paper we propose a novel deep network for biomechanical simulation of brain atrophy, and seek to model differential patterns of atrophy following healthy ageing or AD. In this way our model parallels a growing body of deep generative, interpretable or explainable models of disease. This includes \cite{bass_image_2018,bass2020icam,baumgartner_visual_2018,lanfredi2020interpretation} which train generative models to deform \cite{lanfredi2020interpretation} and/or change the appearance \cite{bass_image_2018,bass2020icam,baumgartner_visual_2018,lanfredi2020interpretation} of images, in such a way that it changes their class. By contrast, deep structural causal models such as \cite{pawlowski_deep_2020}, go further to support counterfactual models of disease progression, by associating demographic and phenotypic variables to imaging data, through variational inference on a causal graph.

One challenge with structural causal models is that they require prior hypothesis of a causal graph, defining the directions of influence of different parameters in the model. In this paper, we therefore take a more explicit approach to explainable modelling, training a hyper-elastic strain simulation of brain growth and atrophy, while 
building an explicit simulation of atrophy for different populations and time windows. This supports subject-specific interventions, simulating projections of brain atrophy, following differing diagnoses.

\section{Methods}

\subsection{Data}

Data used in this study were obtained from the Alzheimer’s Disease
Neuroimaging Initiative (ADNI) database\footnote{\url{http://adni.loni.usc.edu/}}. A total of 1054 longitudinal MRI scans, collated from the ADNI1, ADNI2, ADNI-GO and ADNI3 studies, were used. All examples have at least 2 different T1-weighted scans, separated by at least 1 year (range 1 - 14 years). Accelerated MRI data was used for the subjects that don't have non-accelerated images for both time-points. The dataset includes 210 subjects diagnosed with AD, 677 subjects with Mild Cognitive Impairment (MCI), 67 subjects with Significant Memory Loss (SMC) and 92 cognitively normal (CN). From this, subjects were separated into 845 training datasets, 104 validation datasets, and 105 test datasets. An equal distribution of the 4 disease classes was ensured in each set.

\subsection{Preprocessing}
MRI images were segmented into cerebrospinal fluid (CSF), white matter (WM), gray matter (GM), deep gray matter (DGM) and cerebellum using NiftySeg \footnote{http://github.com/KCL-BMEIS/NiftySeg}. The images were parcellated into 138 regions (NeuroMorph parcellations) using the geodesic information flow (GIF) algorithm \cite{cardoso_gif}. We then generate a less granular parcellation of 27 regions that includes the separate cortical lobes, ventricular system and hippocampus, which we use in the model and for our analysis. T1 images were skull stripped based on the segmentations, then resampled to MNI space with rigid registration using FSL's FLIRT \cite{jenkinson2001global}. Data were normalised into the range 0-1 using histogram normalization, based on data from a target subset of 50 subjects.

\subsection{Model overview}

\begin{figure}[t]
  \centering
  \includegraphics[width=\textwidth]{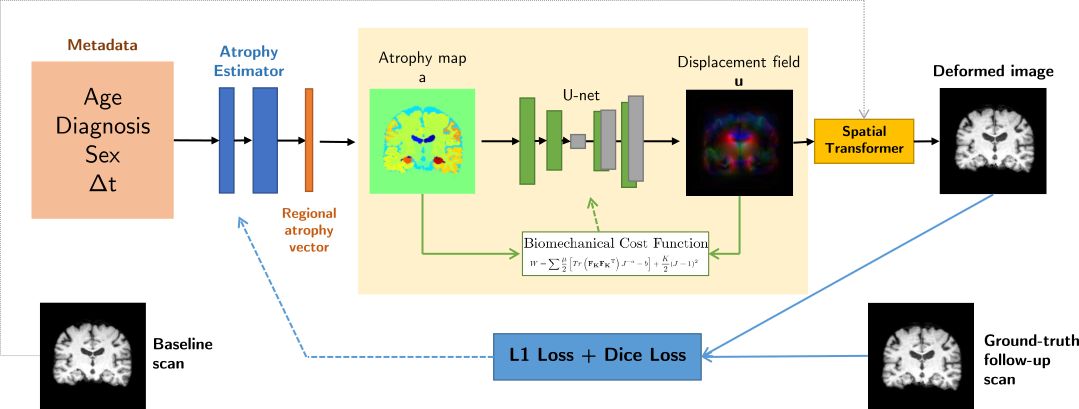}
  \caption{Model architecture: the biomechanical model estimates a deformation field from a prescribed atrophy map corresponding to local volume changes; the atrophy estimator predicts region-wise atrophy values based on demographics and time. In this work, we train the two networks in different stages: the model highlighted in yellow is pre-trained based on a biomechanical cost function (green arrows); the atrophy estimator is trained based on the similarity metrics between simulated and true follow-up image (blue arrows). At inference time, the model estimates a follow-up scan based on metadata and a baseline image alone (the testing procedure can be identified by the black and grey arrows). }
  \label{fig:model}

\end{figure}
Figure \ref{fig:model} offers an overview of the model and the training procedure. The model consists of two networks: an Atrophy Estimator and a Biomechanical Network.

\paragraph{\textbf{The Atrophy Estimator}} is a two-hidden layer (32 and 64 nodes) multi-layer perceptron (MLP) which takes as input 4 demographic variables: biological age (at the time of the first scan - normalized), sex, disease class (CN, SMC, MCI or AD) and time-interval between scans ($\Delta$t). It predicts as output a tensor of size 27, which corresponds to a predicted atrophy or growth value for each region of the brain (using the less granular parcellations, in order to reduce computational cost). The goal of this network is to estimate region-wise values of atrophy and growth, between any 2 longitudinal scans. This vector is then mapped back onto the label image, in order to generate a 3D volumetric map of prescribed atrophies, piece-wise constant, across regions.

\paragraph{\textbf{Biomechanical Network:}}  The goal of the Biomechanical Network is to estimate a displacement field {\bf{u}} from atrophy values, ${a}$, corresponding to local volume changes. In this paper, {\bf{u}} was estimated from a U-net architecture (implemented as for VoxelMorph \cite{balakrishnan_voxelmorph_2019}) and then was used to simulate follow-up scans $X$, from a baseline scan $x$, as $X = x + \mathbf{u}$. A Spatial Transformer was used to apply the deformation field to the original grid and compute the deformed image.

In training, network parameters were optimized based on a biomechanics-inspired cost function.
Following the convention used in modelling growth of biological tissues \cite{rodriguez_stress-dependent_1994,young_automatic_2010}, we model the brain as a Neo-Hookean material and minimise the strain energy density, W:
\begin{equation}
W=\sum{\frac{\mu}{2}\left[Tr\left(\bf{F_K} \bf{F_K}^{\rm{T}}\right) J^{-2/3}-3\right]+\frac{K}{2}(J-1)^{2}}
\end{equation}
Here, \(J = \rm{det}(\bf{F_K}) \), and the elastic deformation $\mathbf{F_K}$ is responsible for driving equilibrium. This is given by $\mathbf{F_K} = \mathbf{F} \cdot \mathbf{G^{-1}}$, where $\mathbf{F} = \nabla \mathbf{u} + \mathbf{I}$ is the total deformation gradient  and $\mathbf{G}$ is the applied growth, $\mathbf{G} = (a^{-1/3})\mathbf{I}$. $a$ represents relative changes in volume, and we assume isotropic growth/atrophy. \( \mu \) is the shear modulus and \( K = 100 \mu\) is the bulk modulus. We define \( \mu = 1 \) for pixels belonging to GM and WM, and set \( \mu = 0.01 \) for the CSF, which we model as a quasi-free tissue. As only the tissues inside the skull suffer deformation, we add a loss term to encourage zero displacement in the voxels outside of the CSF. We also minimize the displacement at the voxel corresponding to the centre of mass of the brain. The total cost function is:
\begin{equation}
\mathcal{L}_{Biomechanical} = W + \lambda_1 \sum{\|\bf{u}_{background}\|}^2 + \lambda_2 \|\bf{u}_{center}\|^2,
\end{equation}
where \( \lambda_1, \lambda_2 \) are hyperparameters weighting the contribution of these terms.

\subsection{Training and Evaluation}

We train the two networks of our model in two separate stages:

\paragraph{\textbf{Pre-training the Biomechanical Model:}} Here, ground-truth per-region atrophy maps were first calculated from the volume ratio between the 2 time-points for each of the original 138 NeuroMorph parcellations \(({a}_{ground-truth} = V_{1}/V_{2})\). These were then used to simulate a range of possible atrophy maps by sampling, for each region, from a uniform distribution of plausible atrophies (with range constrained between the min and max values of each population). In this way, the diversity of training samples seen by the model was increased.

At each iteration, the model was trained with either a subject-specific ground-truth atrophy or an atrophy pattern randomly sampled from these distributions. We note that the aim here is to train the network to estimate displacement fields from any reasonable value of prescribed growth or atrophy, rather than learn deformation patterns from the population.
The biomechanical model was trained for 200 epochs using a mini-batch size of 6 and ADAM optimizer with a learning rate of \( 1\times10^{-4}\). Based on our previous experiments, we set \( \lambda_1 = 10^{-1} \) and $\lambda_2 = 10^{2}$.

\paragraph{\textbf{Atrophy simulation:}} Subsequently, the atrophy estimator was trained to predict the atrophies from the subject demographics and time-window. To train this network, the MLP outputs are applied to the pre-trained biomechanical model to compute the corresponding displacement field, simulated image and simulated parcellations. We then update the weights of the MLP based on the average Soft Dice Loss across the 27 parcellations and the $\mathcal{L}_{1}$ loss between simulated follow-up scan and ground-truth follow-up scan. The total loss of this network is therefore given by:
\begin{equation}
\mathcal{L}_{MLP} = Soft Dice + 0.1 \mathcal{L}_{1}
\end{equation}
The network was trained for 50 epochs, with batch size= 3 and learning rate = \( 1\times10^{-4}\).

\section{Experimental Methods and Results}

\subsection{Evaluation of Biomechanical Model}

\begin{figure}[b]
  \centering
  \includegraphics[width=0.93\textwidth]{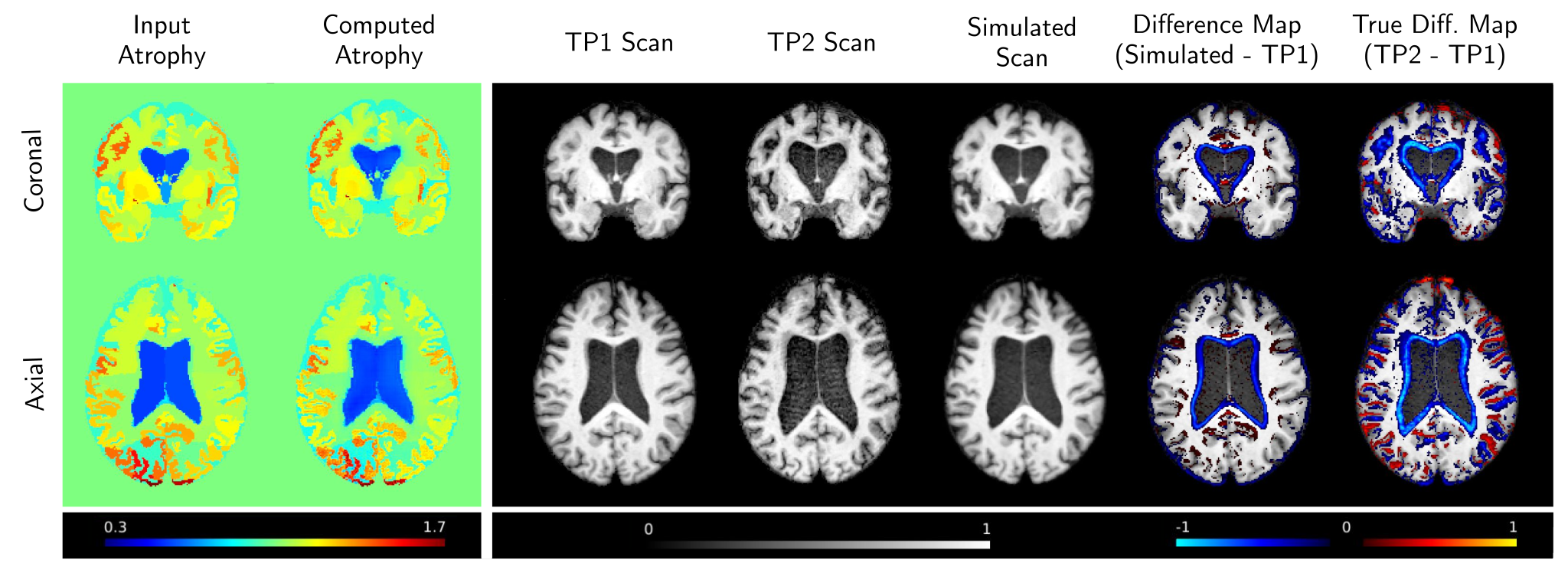}
  \caption{Results of biomechanical model applied to a ground-truth atrophy map. The simulated follow-up scan shows atrophy of the ventricles and cortex that approximates the true difference map between the scans. }
  \label{fig:results1}
\end{figure}

We apply the trained biomechanical model to the region-wise ground-truth volume change values of the 105 subjects of the test set. Figure \ref{fig:results1} shows a representative example of a prescribed atrophy map, computed atrophy (det($\mathbf{F}$)), simulated follow-up scan and corresponding ground-truth for a subject diagnosed with MCI, with a time-span between scans of 7 years.

We evaluate the performance of the network by comparing the prescribed atrophy maps with the computed atrophy using the Mean Squared Error (MSE), and compare the simulated images and segmentations to the ground-truth using MSE and dice overlap scores. In addition, we calculate the Absolute Symmetric Percentage Volume Change (ASPVC) between the simulated and ground-truth follow up images as in \cite{khanal_biophysical_2016}. The objective is to show that the biomechanical network can estimate realistic deformation fields when provided with a specific atrophy map corresponding to local volume changes. 

\newpage
Table \ref{table1} shows the evaluation metrics over the test set, with focused analysis of the dice and ASPVC metrics for ventricles and cortex. We report high dice overlap and low ASPVC for the ventricular region across all subjects, showing that the model can accurately simulate the deformation patterns in the ventricles. Calculated ASPVC values are inside the range (2\% - 5\%) of values reported in \cite{khanal_biophysical_2016}, which simulated atrophy using a FDM model. Lower values of dice overlap for the cortex region can be explained partially by registration differences between the two scans, which influence not only the comparison between simulated and ground-truth image, but also the "ground-truth" volume changes used as input to the model, that are calculated from the parcellations. Note throughout, that 0\% volume change could only be expected for the images if the model was prescribed precise voxel-wise atrophy values.

\begin{table}[t]
  \caption{Evaluation metrics (Mean and Standard Deviation) calculated over the 105 subjects of the test set, for the biomechanical model applied to the ground-truth atrophy maps.}
  \label{table1}
  \centering
  \small
  \begin{tabular}{p{20mm}ccccccc}
    \toprule
     & MSE$_{atrophy}$ & MSE$_{Image}$ & Dice$_{vent}$  & Dice$_{cortex}$ & ASPVC$_{vent}$ & ASPVC$_{cortex}$\\
    \midrule
    \centering\textbf{Mean} & \( 1.27\times10^{-4}\) & \( 2.38\times10^{-3}\) & 0.901  & 0.760 & 2.6 \% & 4.1 \%  \\
    \makecell{ \textbf{Standard} \\ \textbf{Deviation}} & \( 2.87\times10^{-4}\) & \(  1.65\times10^{-3}\) &  0.047 &  0.065 &   4.5 \% &   5.0 \% \\
    \bottomrule
  \end{tabular}
\end{table}

\subsection{Evaluation of atrophy estimation}

Our next aim is to use this model to simulate patterns of atrophy according to different conditions, including the elapsed time between scans and the disease status. In this section, we show that: 1) our atrophy estimation model is capable of simulating follow-up scans consistent with the ground-truth; 2) the model can differentiate between atrophy for healthy, MCI and AD subjects; 3) the model can project forward in time.  

\begin{figure}[h]
  \centering
  \includegraphics[width=0.9\textwidth]{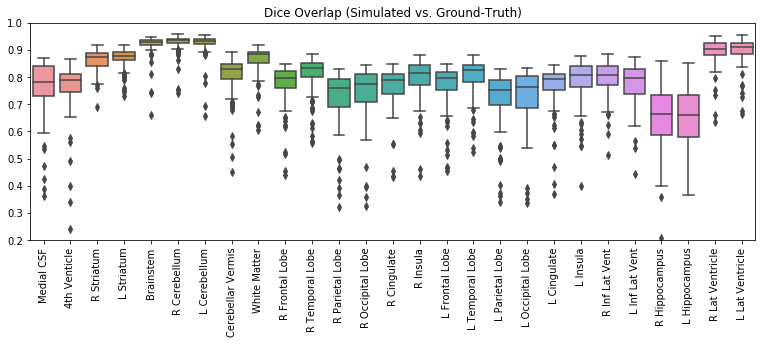}
  \caption{Dice overlap scores between the simulated and ground-truth parcellations for each of the 27 regions, computed across the 105 subjects of the test set.}
  \label{fig:results2}
\end{figure} 

\subsubsection{Comparison with ground-truth:}

We start by evaluating our atrophy estimation model on the 105 subjects of the test set and compare the simulated follow-up scans with the ground-truth images. This evaluation is done in a similar manner to Section 3.1, but here using the full model with the atrophy maps, $a$, estimated from the metadata using the MLP atrophy simulator. Figure \ref{fig:results2} shows the dice overlap between simulated and ground-truth follow-ups, over all considered regions. 

\subsubsection{Predicting trajectories of disease:}

\begin{figure}[h]%
    \centering
    \subfloat[\centering ]{{\includegraphics[width=0.45\textwidth]{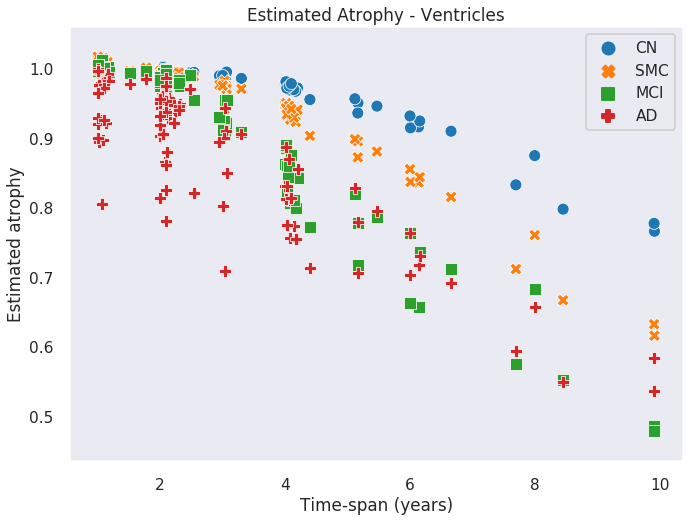} }}%
    \qquad
    \subfloat[\centering ]{{\includegraphics[width=0.45\textwidth]{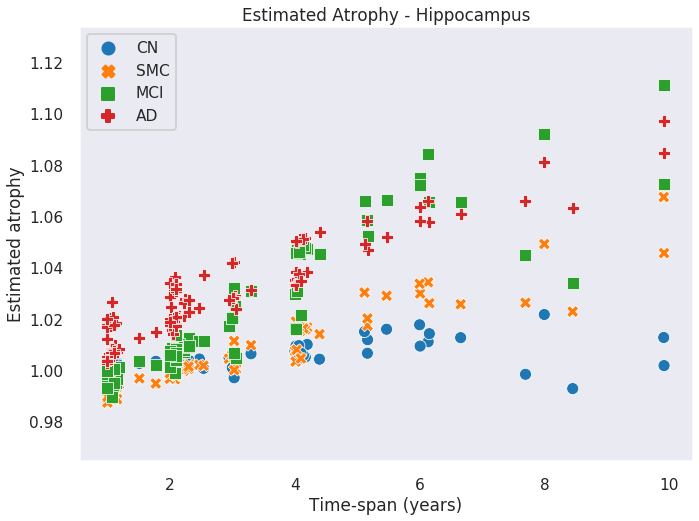} }}%
    \caption{Predicted atrophy values on (a) ventricles and (b) hippocampus when intervening on disease status. Values of $a < 1$ correspond to regional expansion and $a > 1$ correspond to shrinking.}%
    \label{fig:results_predatr}%
\end{figure}

In order to evaluate the ability of our model to differentiate between healthy aging, MCI and AD, we use our trained MLP to estimate atrophy patterns for the different classes. For this, we use the metadata from the 105 subjects of the test set and re-estimate the atrophy maps by intervening on input channel corresponding to the diagnosis class. We therefore calculate 4 atrophy maps for each subject (disease class = \{CN, SMC, MCI, AD\}), and keep the remaining metadata (age, sex, $\Delta t$) as the true values. Figure \ref{fig:results_predatr} shows the predicted atrophy values for the ventricles and hippocampus when intervening on the disease class. 

We performed one-sided paired t-test analysis on the computed atrophies, and conclude that, for the ventricles, the model is able to predict statistically significant differences in atrophy distributions between all 4 diagnosis classes (\(P < .001\)). For the hippocampus, the model estimates atrophies that are significantly different when comparing CN vs MCI, CN vs. AD and MCI vs. AD (\(P < .001\)); the distributions for CN and SMC are not significantly different (\(P = 0.89\)).

\begin{figure}[t]
  \centering
  \includegraphics[width=0.6\textwidth]{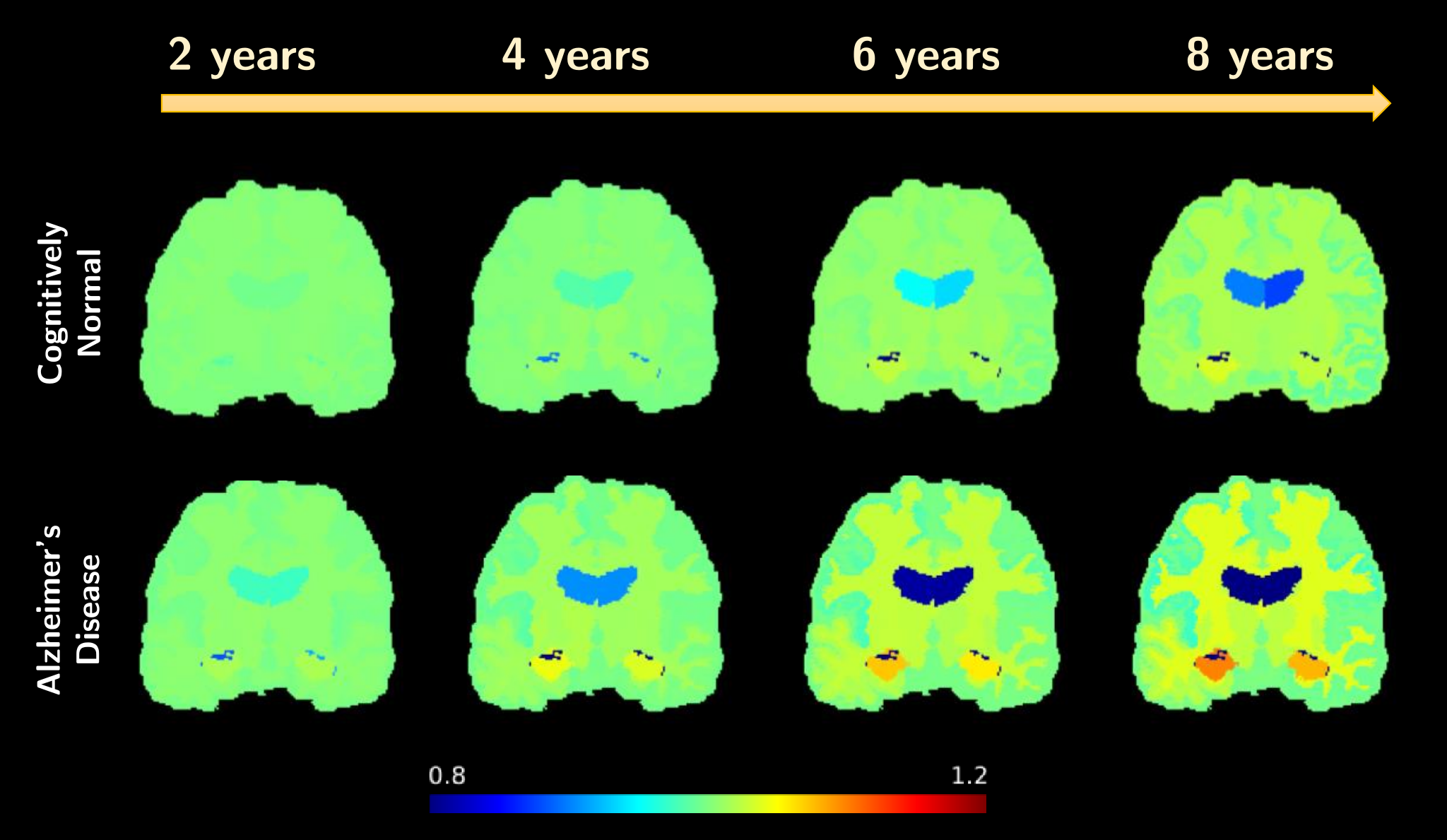}
  \caption{Predicted atrophy trajectory for a subject for healthy ageing (true class) and in the presence of AD. Age at baseline = 82 years.}
  \label{fig:results_atr}
  \vspace{-10pt}
\end{figure}

Finally, to show that the model can predict forward in time, we estimate the atrophy for the subjects of the test set for multiple time-spans ($\Delta$ t = 2, 4, 6, and 8 years). Figure \ref{fig:results_atr} shows the computed atrophy progression when considering healthy aging, and when changing the input class to Alzheimer's Disease. Comparing the trajectories for both cases, it is visible that the model predicts, as expected, larger values of atrophy across the brain tissue, including the ventricles and hippocampus.

\section{Discussion and Future Work}

The results presented here show that the proposed framework can be used to simulate structural changes in brain shape resulting from neurodegenerative disease, and differentiate between healthy and diseased atrophy patterns. In the present case, atrophy was encoded to predict only demographic trends; however our goal is to expand the atrophy estimator network to better model subject-specific heterogeneity by considering information present in the baseline image.

By estimating atrophy values from a small number of variables, the current framework can be used as a simple simulator of disease progression where one can easily intervene on these inputs, including disease status, by simply changing the class. However, while the model can clearly differentiate between healthy and AD subjects, it is well documented that differentiation between MCI and AD is a complex task due to the heterogeneous nature of these disorders. This is reflected in the results from Figure \ref{fig:results_predatr}, and in particular for larger time-windows between scans, for which there are fewer data for the AD class. In the future, and in addition to exploring the use of imaging data as input to the atrophy estimator, we aim to include other metrics of cognitive assessment as input to the network, such as the Mini-Mental State Examination (MMSE) and Alzheimer’s Disease Assessment Scale-Cognition (ADAS-Cog) in order to more accurately predict disease trajectories. We also aim to evaluate the impact of class imbalance on the network training, and address this by including more data from healthy subjects, as well as exploring techniques of oversampling and weighted loss when training the network. 

In this work, we estimate region-wise atrophy maps, which are then used as input to the biomechanical model. In future work, and in order to model patient-specific trajectories of disease, we plan on using the region-wise atrophy estimates as priors to further compute subject-specific voxel-wise patterns that more accurately represent true atrophy patterns. 

Note, although we focus on modelling brain atrophy with age, this proposed model can be translated to other tasks, including brain growth, and can support the use of different biomechanical models of tissue deformation.

\section*{Acknowledgements}

The data used in this work was funded by the Alzheimer's Disease Neuroimaging Initiative (ADNI) (National Institutes of Health Grant U01 AG024904) and DOD ADNI (Department of Defense award number W81XWH-12-2-0012).

\bibliographystyle{splncs04}

\bibliography{paper32.bib}
\end{document}